\documentclass{jpsj3}
\usepackage{graphicx}

\begin{document}

\title{Non-equilibrium Relations for Spin Glasses with Gauge Symmetry}
\author{Masayuki Ohzeki and Hidetoshi Nishimori}
\inst{Department of Physics, Tokyo Institute of Technology, Oh-okayama, Meguro-ku,
Tokyo 152-8551, Japan}
\abst{
We study the applications of non-equilibrium relations such as the Jarzynski
equality and fluctuation theorem to spin glasses with gauge symmetry.
It is shown that the exponentiated free-energy difference appearing in
the Jarzynski equality reduces to a simple analytic function written
explicitly in terms of the initial and final temperatures if the
temperature satisfies a certain condition related to gauge symmetry.
This result is used to derive a lower bound on the work done during the
non-equilibrium process of temperature change.
We also prove identities relating equilibrium and non-equilibrium
quantities.
These identities suggest a method to evaluate
equilibrium quantities from non-equilibrium computations, which may
be useful to avoid the problem of slow relaxation in spin glasses.
}
\maketitle

\section{Introduction}

Equilibrium and non-equilibrium properties of spin glasses have been studied
for many years by experimental, numerical and analytical methods \cite{Rev1,Rev2,Rev3}. 
Although most of the theoretical problems have been solved
fairly satisfactorily at the mean-field level \cite{MPV}, it is still
difficult to establish analytical results for finite-dimensional systems.
Numerical approaches are powerful tools in finite dimensions but are often
hampered by slow relaxation if one wishes to evaluate equilibrium quantities
at low temperatures, and ingenious methods have been proposed to circumvent
this difficulty \cite{Cl,EX,NJ,Iba,Pop1,NE}. In particular, Neal \cite{NJ}
proposed annealed importance sampling, in which one changes the temperature
and measures physical quantities in a spirit similar to the Jarzynski
equality \cite{J1,J2}. Hukushima and Iba \cite{Iba,Pop1} improved Neal's
method by incorporating a branching process for the stability of the
algorithm. Their method, which they called population annealing, showed
outstanding performance comparable to the exchange Monte Carlo.

Non-equilibrium relations such as the Jarzynski equality and fluctuation
theorem \cite{J3,J4,J5} represent important developments in non-equilibrium
statistical physics because they directly relate non-equilibrium and
equilibrium quantities (Jarzynski equality) or the probability of a
non-equilibrium process and its inverse process (fluctuation theorem). It is
the purpose of the present paper to apply the non-equilibrium relations to
the context of spin glasses and show that non-trivial simplifications are
observed under certain conditions on the system parameters. Several
additional non-equilibrium relations are also derived that connect
equilibrium and non-equilibrium quantities. These results are not only
interesting in their own right but may be useful to extract information from
non-equilibrium numerical simulations following the idea of Neal and
Hukushima and Iba.

This paper is organized as follows. In the next section, we recall the basic
formulation in order to fix the notation and set the stage for further
developments in the following sections. In \S 3, we establish several
non-equilibrium relations by using gauge symmetry. In the last section,
we give a summary of the results obtained in the present study.

\section{Formulation}

Let us consider the $\pm J$ Ising model of spin glasses on an arbitrary
lattice, 
\begin{equation}
H=-\sum_{\langle ij\rangle }J_{ij}S_{i}S_{j},  \label{H1}
\end{equation}
where the distribution function of quenched randomness is specified as 
\begin{eqnarray}
P(\tau _{ij}) =p\delta (\tau _{ij}-1)+(1-p)\delta (\tau _{ij}+1) =\frac{%
\mathrm{e}^{K_{p}\tau _{ij}}}{2\cosh K_{p}}  \label{P2}
\end{eqnarray}
with $\tau_{ij}$ being the sign of $J_{ij}$ ($J_{ij}=\tau_{ij}J$). The
parameter $K_p$ has been defined as $\mathrm{e}^{-2K_p}=(1-p)/p$. The product $\beta J
$ will be written as $K$. The following analyses can readily be applied to
other distribution functions of $J_{ij}$ as long as they satisfy a certain
type of gauge symmetry \cite{HNbook,HN81}.

Suppose that the system evolves following a stochastic dynamics governed by
the master equation. For simplicity, we will formulate our theory for
discrete time steps although the continuous case can be treated similarly.
We change the value of the coupling $K$ from $K_0$ at $t=t_0$ to $K_T$ at $%
t=t_n$ in $n$ steps of time evolution, ($K_0, K_1, \cdots, K_n=K_T$).
Correspondingly, the spin configuration changes as $\mathbf{S}_0$ (at $t=t_0$%
), $\mathbf{S}_1$ (at $t=t_1$), $\cdots, \mathbf{S}_n$ (at $t=t_n$). These
configurations will be collectively denoted as $\{ \mathbf{S}\}$. Notice
that each $\mathbf{S}_i$ stands for a configuration of $N$ spins at time $%
t=t_i$. The system is assumed to be in equilibrium at $t=t_0$.

The fluctuation theorem \cite{J3,J4,J5} relates the probability $P_{K_{0}
\rightarrow K_{T}}(\{\mathbf{S}\})$ that such a sequence is realized with
the probability $P_{K_{T} \rightarrow K_{0}}(\{\mathbf{S}\})$ of the inverse
process as 
\begin{equation}
\frac{P_{K_{0} \rightarrow K_{T}}(\{\mathbf{S}\})}{P_{K_{T} \rightarrow
K_{0}}(\{\mathbf{S}\})} \mathrm{e}^{-\beta W } = \frac{Z(K_{T};\{\tau_{ij}\}
)}{Z(K_{0};\{\tau_{ij}\} )}.  \label{F-theorem}
\end{equation}
Here $W$ is the work done to the system during the non-equilibrium process.
More precisely, the work for a single time step $\delta W$ is defined as 
\begin{equation}
\beta \cdot \delta W=( K_{i+1}-K_i)E(\mathbf{S}_i).
\end{equation}
where $E(\mathbf{S}_i)$ is the instantaneous energy for the spin configurations $\mathbf{S}_i$ given by the Hamiltonian (\ref{H1}) divided by $J$.
The right-hand side of eq. (\ref{F-theorem}) is the ratio of equilibrium
partition functions at two different couplings with a fixed configuration of
quenched randomness.

Equation (\ref{F-theorem}) immediately leads to, for an observable $O(\{ 
\mathbf{S}\})$, 
\begin{eqnarray}
\left\langle O(\{\mathbf{S}\})\mathrm{e}^{-\beta W }\right\rangle _{K_{0}\rightarrow
K_{T}} =\langle O_{\mathrm{r}}(\{\mathbf{S}\})\rangle _{K_{T}\rightarrow
K_{0}}\frac{Z(K_{T};\{\tau_{ij}\} )}{Z(K_{0};\{\tau_{ij}\} )},  \label{JE2}
\end{eqnarray}
where $O_{\mathrm{r}}$ denotes the observable which depends on the backward
process $K_{T}\rightarrow K_{0}$. The brackets with subscript $K_0\to K_T$
denote the average with the weight $P_{K_{0} \rightarrow K_{T}}(\{\mathbf{S}%
\})$ over possible non-equilibrium processes. For $O=O_{\mathrm{r}}=1$, eq. (%
\ref{JE2}) reduces to the Jarzynski equality\cite{J1,J2}.

If we choose an observable depending only on the final state, which we
denote by $O_{T}$, instead of $O(\{\mathbf{S}\})$, $O_{\mathrm{r}}$ becomes
an observable at the initial state in the backward process. Then $\langle O_{%
\mathrm{r}}\rangle_{K_{T}\rightarrow K_{0}}$ equals to the ordinary thermal
average at the initial equilibrium state with the coupling constant $K_T$,
to be denoted by $\langle \cdots \rangle_{K_T}$, and eq. (\ref{JE2}) reads 
\begin{equation}
\left\langle O_{T}\mathrm{e}^{-\beta W}\right\rangle _{K_{0}\rightarrow
K_{T}}=\langle O\rangle _{K_{T}}\frac{Z(K_{T};\{\tau_{ij}\})}{%
Z(K_{0};\{\tau_{ij}\})}.  \label{JE3}
\end{equation}


\section{Non-equilibrium relations on the Nishimori Line}

We now consider the application of the relations (\ref{JE2}) and (\ref{JE3})
to the context of spin glasses.

\subsection{Gauge-invariant quantities}

We apply the non-equilibrium relation (\ref{JE3}) to a gauge-invariant
quantity $G(\{\tau_{ij}\})$. After the configurational average (to be
expressed as $[\cdots ]_{K_p}$), we have 
\begin{eqnarray}
\left[ \left\langle G_{T}(\{\tau_{ij}\})\mathrm{e}^{-\beta
W}\right\rangle_{K_0\rightarrow K_{T}}\right]_{K_{p}} =\left[ \langle
G(\{\tau_{ij}\})\rangle_{K_{T}}\frac{Z(K_{T};\{\tau_{ij}\})}{%
Z(K_{0};\{\tau_{ij}\})}\right] _{K_{p}}.  \label{GI1}
\end{eqnarray}
The quantity on the left-hand side is the configurational as well as
non-equilibrium averages of the observable $G(\{\tau_{ij}\})$ at final time $%
T$, that is after the protocol $K_{0}\rightarrow K_{T}$ with the factor $%
e^{-\beta W}$. On the other hand, $\langle G(\{\tau_{ij}\})\rangle_{K_{T}}$
on the right-hand side means the configurational and thermal average of the
equilibrium state for the final Hamiltonian.

Let us apply the gauge transformation $S_i\to S_i\sigma_i,~J_{ij}\to
J_{ij}\sigma_i\sigma_j~(\forall i, j)$ \cite{HNbook,HN81}. The right-hand
side of eq. (\ref{GI1}) is then rewritten explicitly as, 
\begin{eqnarray}
\left[ \langle G(\{\tau_{ij}\})\rangle _{K_{T}}\frac{Z(K_{T};\{\tau_{ij}\})}{%
Z(K_{0};\{\tau_{ij}\})}\right]_{K_{p}} =\sum_{\{\tau_{ij}\} }\frac{\langle
G(\{\tau_{ij}\})\rangle _{K_{T}}\prod_{\langle ij\rangle }\mathrm{e}^{K_p
\tau_{ij}\sigma_i \sigma_j}}{(2 \cosh K_p)^{N_B}}\frac{Z(K_{T};\{\tau_{ij}\})%
}{Z(K_{0};\{\tau_{ij}\})},
\end{eqnarray}
where $N_B$ is the number of bonds on the lattice. All the quantities in
this equation are invariant under the gauge transformation. After the
summation over $\{\sigma_i\}$ and division by $2^N$, we obtain \cite%
{HNbook,HN81} 
\begin{eqnarray}  \label{JENL1}
\left[ \langle G(\{\tau_{ij}\})\rangle _{K_{T}}\frac{Z(K_{T};\{\tau_{ij}\})}{%
Z(K_{0};\{\tau_{ij}\})}\right]_{K_{p}} =\sum_{\{\tau_{ij}\} }\frac{\langle
G(\{\tau_{ij}\})\rangle _{K_{T}}Z(K_{p};\{\tau_{ij}\})}{2^N(2 \cosh
K_p)^{N_B} }\frac{ Z(K_{T};\{\tau_{ij}\})}{Z(K_{0};\{\tau_{ij}\})}.
\label{JE02}
\end{eqnarray}
It is useful to analyze here the quantity $\left[ \langle
G(\{\tau_{ij}\})\rangle_{K_{T}}\right]_{K_{p}}$. Similarly to the above
calculation, the following identity can be derived by the gauge
transformation, 
\begin{eqnarray}
\left[ \langle G(\{\tau_{ij}\})\rangle _{K_{T}}\right] _{K_{p}}
=\sum_{\{\tau_{ij}\} }\frac{\langle G(\{\tau_{ij}\})\rangle
_{K_{T}}Z(K_{p};\{\tau_{ij}\})}{2^N(2 \cosh K_p)^{N_B} }.  \label{JENL2}
\end{eqnarray}%
Setting $K_{p}=K_{0}$ in eq. (\ref{JENL1}) and $K_{p} = K_{T}$ in eq. (\ref%
{JENL2}), we reach the following non-equilibrium relation, 
\begin{eqnarray}
\left[ \langle G_T(\{\tau_{ij}\})\mathrm{e}^{-\beta W}\rangle
_{K_0\rightarrow K_{T}}\right]_{K_{0}} =\left[ \langle
G(\{\tau_{ij}\})\rangle _{K_{T}}\right]_{K_{T}}\left(\frac{2\cosh K_T}{2
\cosh K_0}\right)^{N_B}.  \label{JENL4}
\end{eqnarray}

If we set $G_{T}(\{\tau _{ij}\})=1$ in eq. (\ref{JENL4}), the Jarzynski
equality for spin glass is obtained, 
\begin{equation}
\left[ \left\langle \mathrm{e}^{-\beta W}\right\rangle _{K_{0}\rightarrow
K_{T}}\right] _{K_{0}}=\left( \frac{2\cosh K_{T}}{2\cosh K_{0}}\right)
^{N_{B}}.  \label{JENL5}
\end{equation}%
Equation (\ref{JENL5}) leads to, using Jensen's inequality for the average
of $\mathrm{e}^{-\beta W}$, 
\begin{equation}
\left[ \langle W\rangle _{K_{0}\rightarrow K_{T}}\right] _{K_{0}}\geq -\frac{%
N_{B}}{\beta }\log \left( \frac{2\cosh K_{T}}{2\cosh K_{0}}\right) .
\end{equation}%
The right-hand side corresponds to $\Delta F$ in the Jarzynski equality in
the usual representation.

By substituting $G_T(\{\tau_{ij}\}) = H$ into eq. (\ref{JENL4}), we obtain 
\begin{eqnarray}
\left[ \left\langle H\mathrm{e}^{-\beta W}\right\rangle _{K_{0}\rightarrow
K_{T}}\right] _{K_{0}} = \left[ \langle H \rangle _{K_{T}}\right]
_{K_{T}}\left(\frac{2\cosh K_T}{2 \cosh K_0}\right)^{N_B}.  \label{Ene}
\end{eqnarray}
This equation shows that the internal energy after the cooling or heating
process starting from a temperature on the Nishimori line (NL) \cite%
{HNbook,HN81}, defined by $K=K_p$ which is in the present case $K=K_p=K_0$,
is proportional to the internal energy in the equilibrium state on the NL
corresponding to the final temperature.

\begin{figure}[tb]
\begin{center}
\includegraphics[width=70mm]{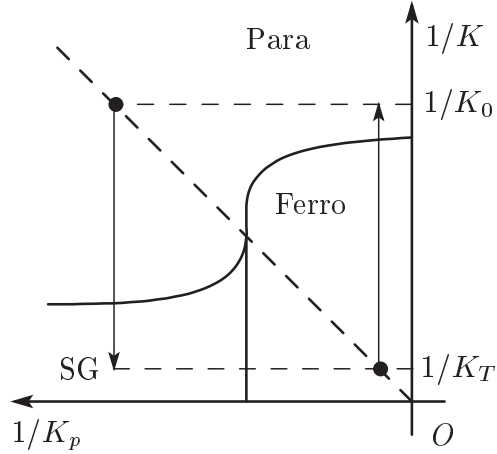}
\end{center}
\caption{{\protect\small The processes of eq. (\protect\ref{NE1}) drawn as
two arrows on the phase diagram. The solid curves are the phase boundaries
and the dashed line of $45^{\mathrm{o}}$ represents the NL.}}
\label{PDNL1}
\end{figure}

It is straightforward to obtain a non-equilibrium relation for
gauge-invariant quantities depending on the intermediate spin
configurations, 
\begin{eqnarray}
\left[ \left\langle G(\{\mathbf{S}\})\mathrm{e}^{-\beta W }\right\rangle
_{K_{0}\rightarrow K_{T}}\right] _{K_{0}} = \left[ \langle G_{\mathrm{r}}(\{%
\mathbf{S}\})\rangle _{K_{T}\rightarrow K_{0}}\right] _{K_{T}} \left(\frac{%
2\cosh K_T}{2 \cosh K_0}\right)^{N_B}.  \label{NEAC}
\end{eqnarray}
For instance, the autocorrelation function satisfies 
\begin{eqnarray}
\left[ \left\langle S_{i}(0)S_{i}(T)\mathrm{e}^{-\beta W}\right\rangle
_{K_{0}\rightarrow K_{T}}\right] _{K_{0}} =\left[ \langle
S_{i}(0)S_{i}(T)\rangle _{K_{T}\rightarrow K_{0}}\right] _{K_{T}}\left( 
\frac{2\cosh K_{T}}{2\cosh K_{0}}\right) ^{N_{B}}.  \label{NE1}
\end{eqnarray}
This gives a non-trivial relation between the cooling and heating processes
but with different amount of quenched randomness characterized by $K_{0}$
and $K_{T}$. Let us consider the cooling process from a temperature on the
NL given by $(1/K_{0},1/K_{0})$ to a point away from the NL $%
(1/K_{0},1/K_{T})$ as depicted by the downward arrow in Fig. \ref{PDNL1}.
The above equation relates this process with the inverse process from a
temperature on the NL $(1/K_{T},1/K_{T})$ to a point away from the NL $%
(1/K_{T},1/K_{0})$ drawn as the upward arrow in Fig. \ref{PDNL1}. The
process of the upward arrow passes through the ferromagnetic and
paramagnetic phases, whereas the cooling process goes through the spin glass
phase. These apparently very different processes are related by eq. (\ref%
{NE1}), which is a non-trivial observation. 

We can establish another type of non-equilibrium relation following Ozeki 
\cite{Ozeki}. We consider a non-equilibrium relaxation of the local
magnetization $\langle S_{i}(T)\rangle _{K_{0}\rightarrow K_{T}}^{\mathrm{F}}
$, where F means that the initial state is $\mathbf{S}_{0}=(+1,+1,\cdots ,+1)
$. We evaluate the time evolution of the local magnetization as 
\begin{eqnarray}
\langle S_{i}(T)\rangle _{K_{0}\rightarrow K_{T}}^{\mathrm{F}} =\sum_{\{%
\mathbf{S}\}}S_{i}(T)\mathrm{e}^{\Delta tM(\mathbf{S}_{n}|\mathbf{S}%
_{n-1};t_{n})} \cdots \mathrm{e}^{\Delta tM(\mathbf{S}_{1}|\mathrm{F}%
;t_{1})},
\end{eqnarray}
where $\Delta t (\ll 1)$ is the unit time interval and $M(\mathbf{S}|\mathbf{%
S}^{\prime };t_{n})$ is the transition rate from state $\mathbf{S}^{\prime }$
to state $\mathbf{S}$ following the master equation. The initial condition F
is different from the case of non-equilibrium relations, where the
equilibrium distribution is assumed initially.

Let us apply the gauge transformation to $\langle S_{i}(T)\rangle
_{K_{0}\rightarrow K_{T}}^{\mathrm{F}}$ \cite{Ozeki}, 
\begin{eqnarray}
\langle S_{i}(T)\rangle _{K_{0}\rightarrow K_{T}}^{\mathrm{F}} =\sum_{\{%
\mathbf{S}\}}S_{i}\sigma _{i}(T)\mathrm{e}^{\Delta tM(\mathbf{S}_{n}|\mathbf{%
S}_{n-1};t_{n})} \cdots \mathrm{e}^{\Delta tM(\mathbf{S}_{1}|\sigma ;t_{1})}.
\end{eqnarray}%
The configurational average for $\langle S_{i}(T)\rangle _{K_{0}\rightarrow
K_{T}}^{\mathrm{F}}$ is thus 
\begin{eqnarray}
&&\left[ \langle S_{i}(T)\rangle _{K_{0}\rightarrow K_{T}}^{\mathrm{F}}%
\right] _{K_{p}}  \nonumber \\
&& \quad=\sum_{\{\tau _{ij}\}}\sum_{\{\mathbf{S}\}}S_{i}(T)\sigma _{i}\frac{1}{(2\cosh K_{p})^{N_{B}}}\prod_{\langle ij\rangle }\mathrm{e}%
^{K_{p}\tau _{ij}\sigma _{i}\sigma _{j}} \mathrm{e}^{\Delta tM(\mathbf{S}%
_{n}|\mathbf{S}_{n-1};t_{n})} \cdots \mathrm{e}^{\Delta tM(\mathbf{S}%
_{1}|\sigma ;t_{1})}.
\end{eqnarray}%
If we set $K_{p}=K_{0}$ and take the summation over all configurations of $%
\sigma $, we obtain 
\begin{eqnarray}
\left[ \langle S_{i}(T)\rangle _{K_{0}\rightarrow K_{T}}^{\mathrm{F}}\right]
_{K_{0}} =\sum_{\{\tau _{ij}\}}\frac{Z(K_{0};\{\tau_{ij}\})}{2^{N}(2\cosh K_{0})^{N_{B}}}%
\langle S_{i}(T)S_{i}(0)\rangle _{K_{0}\rightarrow K_{T}} .  \label{JE17}
\end{eqnarray}%
Since the autocorrelation function is gauge-invariant, the right-hand side
can be shown to be the configurational average using the same method as in
eq. (\ref{JENL2})\cite{Ozeki}, 
\begin{equation}
\left[ \langle S_{i}(T)\rangle _{K_{0}\rightarrow K_{T}}^{\mathrm{F}}\right]
_{K_{0}}=\left[ \langle S_{i}(0)S_{i}(T)\rangle _{K_{0}\rightarrow K_{T}}%
\right] _{K_{0}}.  \label{JE18}
\end{equation}%
Similarly we can prove that the autocorrelation function with the
exponentiated work satisfies 
\begin{eqnarray}
\left[ \langle S_{i}(T)\mathrm{e}^{-\beta W}\rangle _{K_{T}\rightarrow
K_{0}}^{\mathrm{F}}\right] _{K_{T}} =\left[ \langle S_{i}(0)S_{i}(T)\mathrm{e%
}^{-\beta W}\rangle _{K_{T}\rightarrow K_{0}}\right] _{K_{T}}.  \label{JE19}
\end{eqnarray}
Comparison of eqs. (\ref{NE1}) (with $K_T$ and $K_0$ exchanged), (\ref{JE18}%
) and (\ref{JE19}) reveals 
\begin{eqnarray}
\left[ \langle S_{i}(T)\rangle _{K_{0}\rightarrow K_{T}}^{\mathrm{F}}\right]
_{K_{0}} = \left[ \langle S_{i}(T)\mathrm{e}^{-\beta W}\rangle
_{K_{T}\rightarrow K_{0}}^{\mathrm{F}}\right] _{K_{T}}\left( \frac{2\cosh
K_{T}}{2\cosh K_{0}}\right) ^{N_{B}}.
\end{eqnarray}

\subsection{Gauge-non-invariant quantities}

\begin{figure}[tb]
\begin{center}
\includegraphics[width=70mm]{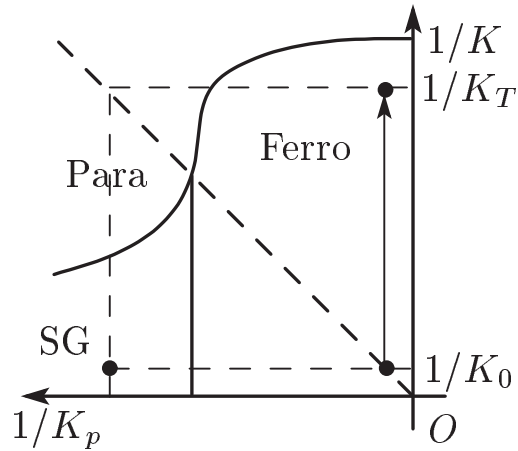}
\end{center}
\caption{{\protect\small Equations (\protect\ref{BJE1}) and (\protect\ref%
{BJE2}) relate equilibrium physical quantities evaluated at a point $(1/K_0, 1/K_T)$
(the lower-left dot in each panel) with other physical quantities evaluated
by non-equilibrium processes shown in an arrow. The lower-left dot is in the
spin glass phase  whereas the corresponding arrow is in the ferromagnetic phase. } }
\label{PDNL2}
\end{figure}

As a typical gauge-non-invariant quantity, we choose $S_i(T)$ for $O$ in eq.
(\ref{JE2}). After the configurational average, we have 
\begin{equation}
\left[ \left\langle S_{i}(T)\mathrm{e}^{-\beta W}\right\rangle
_{K_{0}\rightarrow K_{T}}\right] _{K_{p}}=\left[ \langle S_{i}\rangle
_{K_{T}}\frac{Z(K_{T};\{\tau _{ij}\})}{Z(K_{0};\{\tau _{ij}\})}\right]
_{K_{p}}.
\end{equation}%
Gauge transformation for the right-hand side in this equation yields 
\begin{eqnarray}
\left[ \langle S_{i}\rangle _{K_{T}}\frac{Z(K_{T};\{\tau _{ij}\})}{%
Z(K_{0};\{\tau _{ij}\})}\right] _{K_{p}} =\sum_{\{\tau _{ij}\}}\frac{%
Z(K_{T};\{\tau _{ij}\})}{Z(K_{0};\{\tau _{ij}\})}\langle S_{i}\rangle
_{K_{T}}\sigma _{i}\prod_{\langle ij\rangle }\frac{\mathrm{e}^{K_{p}\tau
_{ij}\sigma _{i}\sigma _{j}}}{2\cosh K_{p}}.
\end{eqnarray}%
As usual, we sum both sides of this equation over all the possible configurations
of $\sigma $ and divide the obtained quantity by $2^{N}$ to find 
\begin{eqnarray}
\left[ \langle S_{i}\rangle _{K_{T}}\frac{Z(K_{T};\{\tau _{ij}\})}{%
Z(K_{0};\{\tau _{ij}\})}\right] _{K_{p}} =\sum_{\{\tau _{ij}\}}\frac{%
Z(K_{p};\{\tau _{ij}\})}{2^{N}(2\cosh K_{p})^{N_{B}}}\langle S_{i}\rangle
_{K_{T}}\langle \sigma _{i}\rangle _{K_{p}}\frac{Z(K_{T};\{\tau _{ij}\})}{%
Z(K_{0};\{\tau _{ij}\})}.  \label{S1}
\end{eqnarray}%
The following relation can also be derived in a similar manner, 
\begin{eqnarray}
\left[ \langle S_{i}\rangle _{K_{0}}\right] _{K_{p}} =\sum_{\{\tau _{ij}\}}%
\frac{Z(K_{p};\{\tau _{ij}\})}{2^{N}(2\cosh K_{p})^{N_{B}}}\langle
S_{i}\rangle _{K_{0}}\langle \sigma _{i}\rangle _{K_{p}}.  \label{S2}
\end{eqnarray}%
Setting $K_{p}=K_{0}$ in eq. (\ref{S1}) and $K_{p}=K_{T}$ in eq. (\ref{S2}),
we find a relation 
\begin{equation}
\left[ \langle S_{i}\rangle _{K_{T}}\frac{Z(K_{T};\{\tau _{ij}\})}{%
Z(K_{0};\{\tau _{ij}\})}\right] _{K_{0}}=\left[ \langle S_{i}\rangle _{K_{0}}%
\right] _{K_{T}}\left( \frac{2\cosh K_{T}}{2\cosh K_{0}}\right) ^{N_{B}}.
\end{equation}%
Thus a non-equilibrium relation results, 
\begin{equation}
\left[ \langle S_{i}(T)\mathrm{e}^{-\beta W}\rangle _{K_{0}\rightarrow K_{T}}%
\right] _{K_{0}}=\left[ \langle S_{i}\rangle _{K_{0}}\right] _{K_{T}}\left( 
\frac{2\cosh K_{T}}{2\cosh K_{0}}\right) ^{N_{B}}.  \label{BJE1}
\end{equation}%
The same method yields 
\begin{eqnarray}
\left[ \langle S_{0}(T)S_{r}(T)\mathrm{e}^{-\beta W}\rangle
_{K_{0}\rightarrow K_{T}}\right] _{K_{0}} =\left[ \langle S_{0}S_{r}\rangle
_{K_{0}}\right] _{K_{T}}\left( \frac{2\cosh K_{T}}{2\cosh K_{0}}\right)
^{N_{B}}.  \label{BJE2}
\end{eqnarray}%
Equations (\ref{BJE1}) and (\ref{BJE2}) relate the equilibrium physical quantities
evaluated away from the NL (the right-hand sides) with other quantities
measured by non-equilibrium processes from a point on the NL to another
point away from the NL (the left-hand sides) as depicted in Fig. \ref{PDNL2}%
. 

\section{Summary}

We have studied applications of non-equilibrium relations, typically the
Jarzynski equality, to the context of spin glasses with gauge symmetry. It
has been shown that the configurational average greatly simplifies a number
of expressions appearing in non-equilibrium relations. In particular, the
right-hand side of the Jarzynski equality, usually written as $\mathrm{e}^{-\beta
\Delta F}$, reduces to a trivial analytic function of the initial and final
temperatures, which has been used to prove a simple lower bound on the work.
Many identities have also been derived for gauge-invariant and
gauge-non-invariant quantities, which relate physical quantities measured at
quite different environments. Most notably, the equilibrium values of the
single-site magnetization and correlation function have been proved to
be proportional to the non-equilibrium values of the corresponding
quantities measured at different parts of the phase diagram. This result may
possibly be useful to numerically evaluate equilibrium physical quantities
in the spin glass phase from non-equilibrium calculations away
from the spin glass phase with the aid of annealed importance sampling or
population annealing method \cite{NJ,Iba,Pop1}.

\begin{acknowledgments}
This work was partially supported by CREST, JST, and by the Grant-in-Aid for Scientific Research on the Priority Area \textquotedblleft Deepening and Expansion of Statistical Mechanical Informatics\textquotedblright\ by the
Ministry of Education, Culture, Sports, Science and Technology.
\end{acknowledgments}

\end{document}